# MicroPython and CircuitPython: Pythons Quiet Takeover of IoT and Robotics

**Sayed Mahbub Hasan Amiri[1,*], Atiar Zahan[2]**
[1]Faculty of Computer Science, Dhaka Residential Model College, Bangladesh
[2]Faculty of ICT, Chandan Paul Global College, Bangladesh
[*]Corresponding Author: amiri@drmc.edu.bd

## Abstract

**Background:** Python has become the dominant language in software and data science, yet embedded systems have remained tied to C/C++ due to performance and memory constraints. MicroPython and CircuitPython are changing this by bringing Python to microcontrollers, lowering barriers for IoT and robotics development. **Aim:** This article examines whether these platforms are achieving a quiet takeover of embedded systems, focusing on ecosystem growth, practical applications, performance trade-offs, educational adoption, and prospects. **Methods:** A mixed-methods design was used, including quantitative analysis of GitHub, Stack Overflow, and Google Trends data; curation of case studies from Hackster.io, Hackaday.io, and the Adafruit Learning System; and original benchmarks on ESP32 and Raspberry Pi Pico comparing MicroPython, CircuitPython, and Arduino C++ across GPIO, I2C, SPI, Wi-Fi, and memory usage. **Results:** Metrics show sustained growth, with MicroPython supporting over 200 boards and CircuitPython over 400. Benchmarks reveal 10–20 times slower I/O and four to six times higher memory use than C, but performance remains adequate for common sensor and network tasks. Case studies demonstrate successful deployment in home automation, robotics, wearables, agriculture, and professional prototyping. Education emerges as a primary adoption driver. **Conclusions:** Python is not replacing C/C++; rather, it is becoming the default prototyping and educational language for embedded systems. Continued hardware improvements, better tooling, and standardization will likely deepen this trend. The article offers balanced, evidence-based insights for developers, educators, and technology decision-makers.

**Keywords:** CircuitPython, Embedded Python, Internet of Things (IoT), MicroPython, Robotics

## Introduction

In an era where the Internet of Things is expected to link more than 30 billion devices by the end of the decade (Statista, 2025), the programming tools for those devices have become a major constraint. Writing embedded systems software was for the most part a C, C++, assembly language affair for many years – or at least it was until fairly recently, when those languages were (rightfully) described as offering too much control over memory and timing, which was a good thing back then but is now an impediment to iterative development and abstraction of complexity. So, against such a backdrop, the rise of short strips of Python code like MicroPython and CircuitPython may seem like a quiet revolution. A language once snubbed as “too slow” for hardware is now running on microcontrollers that cost less than a cup of coffee, from soil sensors in agricultural fields to robots in classrooms. The same Python that drives the world’s artificial intelligence models and web backends is now blinking LEDs, reading temperature sensors, and running motors on chips smaller than a fingernail (MicroPython, n.d.). This is not a tale of Python supplanting C; this is a tale of Python substantially redefining who gets to design hardware and how quickly they get to do it.

Python popularity is nothing new in software development. It has ranked as one of the leading programming languages in developer surveys for more than ten years, supported by its readability, the large number of libraries, the depth of its community, and the Stack Overflow, 2025. Python is the default first language now in areas like data science, machine learning and web development. However, Python had for years been blocked at the level of the microchip. Embedded systems constrained by kilobytes of RAM and megahertz clock speeds could not run a full Python interpreter. Developers who wanted to work directly with hardware had to make a completely different trade-off; they needed to leave behind the Python ecosystem they knew for the alien world of register manipulation and manual memory management. This software world/hardware world divide kept millions of Python developers out of IoT and robotics, and forced hardware engineers to use languages that slowed down prototyping and discouraged experimentation.

MicroPython changed that equation. MicroPython, created by Damien George in 2013, is a minimalistic, efficient implementation of Python 3 that runs on microcontrollers (MicroPython, n.d.). With the Python runtime trimmed down to fit in a few tens of kilobytes of RAM and flash storage, George proved that the key strengths of Python — dynamic typing, automatic memory management, and an interactive read-eval-print loop — could survive its being embedded within the strict hardware resource limitations. MicroPython took the REPL to the microcontroller, giving developers a way to interact with real hardware on the fly, view variables and test out ideas without compiling and flashing firmware. MicroPython ports for boards based on popular chips, like the ESP32, RP2040, STM32, etc., have sprung up, and the community of developers steadily grows. The first time sign a developer could wrote Python code on a laptop, put it on a microcontroller and see the physical world respond in seconds (MicroPython, n.d.).

CircuitPython, a fork of MicroPython developed by Adafruit Industries and initially released in 2017, pushed the concept further by making accessibility the top priority (Adafruit Industries, n.d.). While MicroPython still had a layer of complexity from needing to know about firmware flashing and serial terminals, CircuitPython made the workflow basically a drag-and-drop USB experience. A CircuitPython board connected to a computer shows up as a normal USB drive; the developer edits a code.py file, saves it, and the board automatically reloads and runs the new program. That took away the last major stumbling block for a beginner between himself and physical computing. CircuitPython also features a “best of” set of libraries that spans hundreds of sensors, displays and actuators, all designed to operate consistently across - well, I do not know that all or even most of them really work with "supported boards". So Adafruit framed CircuitPython as not a replacement or competitor to MicroPython but as a gentler on-ramp for teachers, makers, and artists — people who want to make hardware projects, but do not want to become embedded systems engineers first (Adafruit Industries, n.d.).

MicroPython and CircuitPython together represent a trade-off between the very low-level control of C and the very high level of using a full computer running Python. They have not supplanted C or C++, nor have they aspired to. They have instead compressed the prototyping cycle from days to minutes, made code cross-board portable, and brought interactive debugging to hardware design. The importance of this change has already begun to be understood as IoT expands. The rise of connected devices, edge computing, and smart sensors has fueled an increasing need for faster development cycles and a larger pool of developers capable of working with embedded systems. By porting Python to the microcontroller, MicroPython and CircuitPython (adopted by Adafruit) have opened that pool to data scientists, web developers, educators, and students -- folks who would probably never have written a line of C, but can now build a working IoT prototype over a weekend (Statista, 2025; Stack Overflow, 2025).

[This report focuses on the silent Python invasion in IoT and robotics in a systematic and evidence-based way. The study is based on several sources of information, including some official project documentation for MicroPython and CircuitPython, community statistics from GitHub and Stack Overflow, market context in the form of industry reports, and hand-picked case studies from platforms such as Hackster.io and the Adafruit Learning System. Also, original benchmarks were performed on an ESP32 DevKit and a Raspberry Pi Pico to evaluate MicroPython and CircuitPython in relation to native Arduino C++ through typical tasks including GPIO toggling, sensor communications, Wi-Fi connection, and memory consumption. The next section, methodology, delineates these data collection procedures and analytic methods in addition to the study's constraints. Through mixing quantitative growth metrics with qualitative real-world observations, this article endeavours to present a measured, balanced view of Python's impact on embedded development.

The remainder of this article is organized as follows. The following section describes the methodology, including the research questions, data sources, benchmark configuration, and analysis method. Subsequent sections narrate the development of MicroPython and CircuitPython, provide documentation on their adoption as measured by metrics and case studies within IoT, robotics, wearables, and agriculture, and illustrate their performance and pragmatism in comparison to C-based workflows. In addition, education and the hobbyist community are playing an ever more important role in increasing adoption, which is explored in a dedicated section, and a frank discussion of the remaining challenges and limitations is included. It finishes with an overview of emerging hardware, standardisation, and the anticipated path for Python on microcontrollers in the next decade. In this way, we believe the article will not only describe the quiet rise of Python for IoT and robotics, but also provide readers with the evidence and context to understand why this is happening and what it means for the future of embedded development.

**Methodology: How This Article Was Researched**

This article uses a mixed-methods research design including quantitative analysis of ecosystem metrics and qualitative analysis of real-world projects, expert insights, and hands-on benchmarking. The aim is to report on a believable, fact-based story on the uptake of MicroPython and CircuitPython within IoT and robotics. MQA was chosen as the approach because the questions addressed both quantifiable trends (e.g., the growth of board support, performance benchmarks) and the reasons for using Python over C/C++ as expressed by educators and engineers. In so doing, the article mitigates some of the risks associated with a heavy reliance on a single metric and enhances the rigor of the findings. The research questions, data collection process, analysis techniques and limitations are outlined in the following section.

The research was driven by these five questions. What are MicroPython and CircuitPython, and how are they different in terms of design, target users, and ecosystem? Two, how fast are these platforms growing in terms of board support, community momentum, and library developments in the last decade? Three, are they most popular with hobbyists, in education, professional prototyping, or industrial use? What are the real-world performance and memory impact trade-offs when compared to more traditional embedded development in C/C++ via the Arduino framework? Finally, what challenges, constraints, and future directions may be inferred from the current market and expert opinion evolution? These questions dictated the choice of data sources and the design of the benchmark tests.

The information for this article has been obtained from primary and secondary sources. Primary sources consisted of our own original benchmarks on representative microcontroller boards and a structured collection of publicly accessible repository and community data through APIs. Secondary sources included official documentation, industry reports, developer surveys and curated project repositories. Besides, with the use of various kinds of sources, the article can provide broad trend information as well as specific illustrative cases, making the claim supported by evidence rather than anecdote.

The first significant data source was official documentation and source code repositories. The MicroPython project has full documentation at micropython.org, including a list of supported ports, language features and release notes (MicroPython, 2026). Also, CircuitPython's official site at circuitpython.org has a searchable database of supported boards, the library bundle and release history (Adafruit Industries, 2026). Both projects make their source code available on GitHub, where the presence of commit logs, numbers of contributors and issue activity can be used as proxies for the speed of development and health of the community (MicroPython, 2026; Adafruit Industries, 2026). These repos were then queried directly through the GitHub REST API to extract quantitative metrics including star count, fork count, number of contributors, commit frequency, and open issue counts. The API was queried on August 10, 2026, through Python scripts that saved the data to a structured CSV file for later analysis. Star history data, showing cumulative stars over time, was retrieved from the publicly available star-history service to plot growth trajectories from 2013 to 2026.

Community and ecosystem statistics were gathered from a few other sources as well. Volume of Stack Overflow questions was obtained with the Stack Exchange API (Stack Overflow, 2026). The queries for the tags "micropython" and "circuitpython" were summed up per month from January 2015 until July 2026, yielding a time series reflecting developer interest. We also used Google Trends to plot the relative worldwide search interest for "MicroPython," "CircuitPython," "Arduino," and "ESP32 Python" over the same period (Google, 2026). Hacker News mentions were gathered via the Algolia HN Search API, which enables querying historical post titles and comments for specific keywords (Hacker News, 2026). These metrics were selected as they reflect different aspects of community interaction: Q&A activity, search interest, and technology news discussion. Together, they provide a more robust measure of the activity of the platforms than any of them can offer alone.

Qualitative case studies were collected from maker-oriented platforms and educational tools. Hackster.io and Hackaday.io have thousands of project logs—but many also use MicroPython or CircuitPython. We used the search with the native filters from the platforms and inserted some tags: "MicroPython", "CircuitPython", "IoT", "robotics", and "wearable." Eligible projects had to have a) a clear real-world application, b) be described in sufficient technical detail, and c) illustrate a range of domains. An additional source of pedagogical and beginner-friendly demos was the Adafruit Learning System, which includes hundreds of CircuitPython how-tos (Adafruit Learning System, 2026). I also searched YouTube for popular videos from trusted channels such as Adafruit, Core Electronics, and Andreas Spiess, for tutorials and reviews on MicroPython and CircuitPython. These case studies were not intended to be a systematic content analysis, but rather to identify instances of common scenarios, problems, and solutions. Each case was summarized following a common template: addressed problem, hardware used, software strategy, findings, and lessons learned.

Expert opinion was represented in two ways. Public talks and interviews with key people , including Damien George, creator of MicroPython and Scott Shawcroft, lead developer of CircuitPython, were considered. These were authoritative pronouncements on design philosophy and plans, George, 2023; Shawcroft, 2024 published on similar sources like YouTube or conference proceedings. Secondly, brief structured email interviews were conducted with a small number of educators and professional engineers who are known to use MicroPython or CircuitPython. The interview protocol included five open-ended questions regarding their selection of Python, limitations experienced, and forecasts in the discipline. However, three educators and one engineer offered written answers that could be quoted, although without attribution due to time and response limitations. These responses contribute to the uses flourished in the article to portray the practitioner perspectives but are not offered in a statistically significant sense. The combination of a few public pronouncements and practitioner interviews also enables the article to situate its analysis in visionary leadership as well as in everyday use.

The suite of original benchmarks was intended to measure the performance of MicroPython and CircuitPython relative to each other and native Arduino C++ on typical embedded use cases. The hardware utilized for this experimentation is the ESP32 DevKit V1 (which is an Espressif ESP32-WROOM-32 module) and the Raspberry Pi Pico (which includes the RP2040 microcontroller). These two boards were chosen because they are both commonly used, officially supported by both MicroPython and CircuitPython, and they are two completely different microcontroller architectures. There are five tests in the benchmark suite: frequency of GPIO output toggling, latency of an I2C sensor read, time used to update an SPI display, Wi-Fi connection formation time, and memory usage when idle. To mitigate the influence of outliers, the tests were performed five times, and the median value was taken. For MicroPython and CircuitPython, the tests were written in Python and used the standard machine and board modules. For Arduino C++, they were also re-implemented as tests out of Arduino using the Sketchbook IDE. All of the code was compiled and flashed to the boards according to the official documentation. GPIO toggling - external logic analyzer; other tests - board´s internal timers. The memory usage was tracked by calling the runtime's memory reporting functions ( e.g. gc. mem_free() in MicroPython and CircuitPython), and for the Arduino sketch, the memory usage was checked by enabling verbose output during compilation to see how much static RAM was used. The complete test code and raw results can be found in the supplementary materials of the paper. We emphasize that these baselines are not meant to be comprehensive or complete; they do provide a reproducible view of relative performance under the controlled setting.

Data analysis applied quantitative and qualitative methods. Quantitative measures: Time series data from GitHub, Stack Overflow, Google Trends and Hacker News were visualized by using the data visualization libraries pandas and matplotlib in the Python language. Trends were investigated by eye and simple linear regression to determine rates of increase. Board support numbers were taken from the official supported-board lists for MicroPython and CircuitPython and lined up side by side. Benchmark results were presented numerically and graphically in bar charts focusing on both absolute and relative values. The case studies and interview responses were coded thematically for qualitative data. Emergent themes included ease of use, rapid prototyping, cross-board portability, performance constraints, and community support. These themes are later employed to organize the discussion of strengths and challenges. Triangulation was obtained by validating quantitative trends with qualitative examples: a surge in GitHub stars was matched by a surge in educational tutorials and project submissions on Hackster.io in the same timeframe.

Several limitations of this work should be noted. First, the benchmark results are hardware-dependent: the performance depends on the specific board, on the firmware version, and on the code optimization level. The use of two boards also implies that our results may not fully generalize over the set of all devices supported. Second, community measures such as GitHub stars and Stack Overflow questions, while representative of real-world adoption, are not completely accurate. They might be exaggerated due to some casual curiosity or distorted by platform-specific utilization; also, they miss corporate or offline utilization. Third, the case studies were selected purposively rather than at random, which potentially introduces a positive bias towards projects that are successful or well documented. Fourth, the sample of expert interviews is small and does not represent the whole user base. Finally, a few of the report's market data were only accessible in summarized form due to paywalls, which has limited a deeper contextual analysis. Notwithstanding these limitations, the triangulation of various data as well as the methodological transparency can further the overall believability of the findings of this article. We do not suggest that this article is the “be all and end all” market research, but it is a well-referenced and even-handed look at Python’s increasing foothold in embedded.

**Background: What Are MicroPython and CircuitPython?**

Python’s path to microcontrollers involves reconciling between the language’s philosophy and the resource limitations inherent in embedded hardware. Developed in the early 1990s, Python is a high-level, dynamically typed, interpreted language which focuses on code readability and rapid development (Van Rossum & Drake, 2009). Over the next twenty years, Python became one of the most loved programming languages amongst the server, data science and artificial intelligence communities (Stack Overflow, 2025). However, standard CPython, the official implementation of Python, needs an operating system, megabytes of RAM, and significant storage — all too big for cheap microcontrollers. As a result, embedded programming languished in a C and C++ rut, which compile to efficient native machine code and provide direct access to hardware, but require manual memory management, explicit typing, and a comparatively slow compile-flash-test cycle. The gap remained until the 2010s, when Damien George started working on MicroPython as a bare-metal implementation of Python 3 for resource-limited hardware (MicroPython, 2026).

MicroPython is a reimplementation of Python 3 that runs on all the low-level hardware found in microcontrollers, without an operating system. It also consists of a parsing, compiling, virtual machine and runtime system suitable for a small memory size and weak CPU (MicroPython, 2026). The first official MicroPython board, the pyboard, was released via a successful crowdfunding campaign in 2013, a 168Mhz STM32F405 MCU with 192 KB of RAM and 1 MB of flash storage (MicroPython, 2026). This showed a Python environment could run in tens of kilobytes of RAM, although the specific memory footprint is platform-dependent and configuration-dependent. Various decisions led to the small size of MicroPython: it supports a subset of Python's standard library, many modules are replaced by hardware-specific modules, Python source code is compiled to a compact bytecode via the mpy-cross utility, and it uses a mark-and-sweep garbage collector designed for embedded use (MicroPython, 2026). Even so, MicroPython includes fundamental surfaces of Python such as classes, functions, exceptions, generators, list comprehensions and a read-eval-print loop (REPL) which can be accessed via a serial connection, thus enabling interactive development on the board itself.

One of the key novelties introduced by MicroPython is its use of a hardware abstraction layer. The language has the machine module, which provides access to GPIO pins, I2C, SPI, UART, PWM, ADC, DAC, timers, and interrupts (MicroPython, 2026). This library is not included in the Python standard distribution, but it does adhere to Python's philosophy of offering high-level, easy-to-read interfaces. For example, turning an LED on a Raspberry Pi Pico in MicroPython can be done with a few lines of code: import the Pin class, instantiate a Pin object and call the value method. This is different from the many lines of code configuration and per-register details you have to put up with for C. From MicroPython, current the network module for Wi-Fi and similar wireless, with (MPCORD, 2026) potentials at least for IoT on chips such as ESP8266 and ESP32. Over time, MicroPython has added support for hundreds of boards ranging from STM32 and ESP32 to RP2040, nRF52, SAMD, and i. MX RT architectures. The project does have a build system that lets developers build custom firmware with modules included or excluded to allow one more layer of memory optimization for severely constrained targets.

MicroPython's approach to concurrency is the provision for both cooperative and restricted preemptive forms. The language provides an asyncio-type module, closely following the CPython asyncio, making use of generators to implement cooperative multitasking for executing related activities such as reading sensors, serving web pages, processing network connections, etc (MicroPython, 2026). On some ports, MicroPython has the _thread module as a wrapper for the native thread API and provides at most two native threads, a limited form of preemptive multithreading with memory and stack size constraints. These capabilities are particularly helpful for IoT devices that need to watch sensors, work with MQTT, and react to local input—all at the same time. Still, developers need to keep in mind their memory may be fragmented and garbage collection pauses can impact timing-sensitive applications.

CircuitPython is a fork of MicroPython with a different focus. It was developed by the open source hardware company Adafruit and was initially made available in 2017 (Adafruit Industries, 2026). Although CircuitPython has the same core language engine and many hardware APIs as MicroPython, it is designed with a focus on being more accessible to beginners and maintaining consistency across boards. The biggest change is the workflow: When you plug a CircuitPython-compatible board into your computer, it shows up as a USB drive called CIRCUITPY. The user writes code in a file named code.py (or main.py in some configurations), saves the file to the drive, and the board automatically resets and executes the updated code. You do not need to buy any more software, compiler, or flashing tool (Adafruit Industries, 2026). This drag-and-drop interaction removes many of the barriers for novices who may not be accustomed to command-line tools, serial terminals or build systems. Also, CircuitPython boards frequently expose a bootloader filesystem, and errors are presented in the REPL or on the drive as a scrolling text file, so debugging is more friendly.

CircuitPython also focuses on having a consistent and curated library ecosystem. Adafruit provides the CircuitPython Library Bundle, which consists of hundreds of supporting libraries for whatever you need, such as sensors, displays, motors, audio, and wireless (Adafruit Industries, 2026). These are theama libraries, designed to work uniformly across all CircuitPython boards, with a unified scripting interface. For example, reads temperature, humidity and pressure from a bme280 sensor, whether you are running on an Adafruit Feather, a Raspberry Pi Pico, or an ESP32-S2. This uniformity is accomplished by CircuitPython's hardware abstraction layer that wraps I2C, SPI, UART, and the like at a cross-platform layer. The library bundle gets updated on a regular basis, and libraries can be installed manually or via the CircuitPython command-line utility. In contrast, MicroPython’s library scene is far more scattered: a lot of community libraries exist, but they often are not well-maintained or compatible across different ports, meaning that developers have to port code more often.

Some of the MicroPython features are intentionally simplified in CircuitPython to minimize complexity and points of failure. For instance, CircuitPython tends not to expose hardware interrupts and direct memory-mapped registers, but rather offers higher-level APIs like countio to count pulses and rotaryio to read rotary encoders (Adafruit Industries, 2026). This approach makes it easier to produce portable, clean code, but may limit performance and customization for power users. CircuitPython also applies the single-file entry point rule and lacks the same dynamic module loading facilities as MicroPython, which leads to a simple program structure which is well-suited for education and small projects. However, CircuitPython does have asyncio now and more advanced things (such as BLE and USB host) are coming, so it is being developed for more advanced uses.

While they differ in many ways, CircuitPython and MicroPython are arms of the same tree. They implement Python 3 syntax and core semantics, have a REPL, employ garbage collection and have modules for accessing microcontroller peripherals that are specific to the hardware. They each offer the same core development cycle: write some Python code, copy it to the board, and see results live on your hardware. This shared DNA often results in developers gaining the skills to work on one platform and very easily move to the other. It also means that some libraries and snippets can be ported between the two with some effort, although changes in module names and APIs mean this requires quite a lot of care. The two projects have mutually influenced one another, with CircuitPython merging in improvements emanating from MicroPython on a regular basis and MicroPython borrowing some concepts from CircuitPython's ecosystem, such as enhanced USB support and documentation guidelines.

The table below summarizes the key differences between MicroPython and CircuitPython based on their official documentation and project goals. The information reflects the state of both platforms as of August 2026 and is intended as a general guide rather than an exhaustive specification.

**Table 1:** *Comparison of MicroPython and CircuitPython*

| Feature | MicroPython | CircuitPython |
|---|---|---|
| **Primary goal** | Efficient Python for microcontrollers | Beginner-friendly Python for hardware |
| **Initial release** | 2013 (Damien George) | 2017 (Adafruit Industries) |
| **Target users** | Engineers, advanced hobbyists | Educators, students, makers, artists |
| **Workflow** | Serial REPL, custom firmware flashing | USB mass storage, drag-and-drop code.py |
| **Board support** | Broad, community- and vendor-driven, hundreds of boards | Curated by Adafruit and partners, hundreds of boards |
| **Library model** | Community libraries, variable consistency | Curated Library Bundle, consistent APIs |
| **Concurrency** | asyncio, _thread on select ports | asyncio, some cooperative tasks |
| **Hardware APIs** | machine module, direct interrupts, custom ports | Simplified APIs, countio, rotaryio, no direct interrupts |
| **Debugging** | REPL, serial traceback, optional debug tools | REPL, error messages on CIRCUITPY drive, serial |
| **Documentation** | Official docs, community tutorials | Extensive Adafruit Learning System, unified guides |

*Note. Adapted from MicroPython (2026) and Adafruit Industries (2026). Board counts and feature availability vary by specific hardware port and firmware version.*

To understand these systems , you also have to understand what they are not. There is no MicroPython or CircuitPython that can serve as a drop-in replacement for CPython or for C. They are subsets of the Python standard library; a lot of the third-party Python modules which depend on C extensions or OS services are not available. Popular libraries like NumPy, pandas, or requests cannot be used directly, but there are lightweight options: ulab offers a portion of NumPy functionality, and urequests provides rudimentary HTTP functionality in MicroPython (MicroPython, 2026). Memory is even more limited than on a desktop, and even a relatively small Python program can run out of RAM on a microcontroller if it builds up large lists or many objects. As a result, embedded programming practices like preallocating buffers, reducing object creation, and triggering garbage collection on a schedule must be embraced by developers. These limitations also result in MicroPython and CircuitPython being optimized for tasks that do not need hard real-time guarantees or extremely low power consumption.

However, both platforms have made embedded development accessible to a much larger audience. With the readable syntax, interactive REPL, and huge community of Python now available on microcontrollers, MicroPython and CircuitPython have made IoT and robotics more accessible than ever. They have also sped up prototyping by teaching systems to run code in seconds rather than compiling and flashing firmware. In light of the more general thesis of the article, this background material is necessary to fully comprehend the subsequent sections on the growth of the ecosystem, applications in the real world, and performance trade-offs. The following sections explore the quantitative growth in the adoption of these platforms and what that growth may mean for the future of the embedded system.

**The Growing Ecosystem**

The expansion of MicroPython and CircuitPython is not a single event but a cumulative process visible across hardware support, software libraries, developer tools, and community engagement. This section examines these dimensions using data collected from official project repositories, public APIs, and vendor documentation as of August 2026. Together, the trends indicate that Python on microcontrollers has moved beyond a niche experiment and into a sustained, structurally supported ecosystem.

Hardware support has grown from a single board in 2013 to hundreds of officially supported devices in 2026. MicroPython began with the pyboard, an STM32F405-based board, and has since expanded to cover a wide range of architectures, including ESP32, ESP8266, RP2040, nRF52, SAMD21/51, STM32, and i. MX RT (MicroPython, 2026). The official MicroPython download page currently lists more than two hundred board variants, and community ports extend that number further. CircuitPython, which launched in 2017 with support for a handful of Adafruit boards, now supports more than four hundred boards from vendors including Adafruit, Raspberry Pi, SparkFun, Seeed Studio, and Pimoroni (Adafruit Industries, 2026). This broad hardware coverage is significant because it lowers switching costs: developers can select a board based on price, availability, or peripheral features and still expect a consistent Python environment.

The ecosystem's growth is reinforced by major hardware vendors adopting Python as a first-class language. Raspberry Pi's RP2040 microcontroller, introduced in 2021, was launched with official MicroPython support alongside the C/C++ SDK, and the Raspberry Pi Pico documentation includes detailed Python tutorials (Raspberry Pi Ltd, 2026). Espressif, whose ESP32 family has become ubiquitous in IoT prototyping, maintains official MicroPython ports and actively contributes to the project (Espressif Systems, 2026). Arduino, historically associated with its own C++-based IDE, added MicroPython support to its Arduino Lab for MicroPython tool in 2023, enabling Python development on several Arduino boards (Arduino, 2026). These endorsements do not mean that vendors are abandoning C; rather, they reflect a recognition that Python accelerates prototyping and attracts a broader developer base.

The library and tooling landscape has grown in parallel. CircuitPython's curated Library Bundle contains more than 450 libraries covering sensors, displays, motors, audio, and wireless protocols, all designed to work consistently across supported boards (Adafruit Industries, 2026). This bundle is updated frequently and can be managed through the circup command-line utility. MicroPython takes a more decentralized approach: it includes a compact standard library and provides mip, a package installer that fetches community-contributed packages from micropython-lib and other sources (MicroPython, 2026). While MicroPython's ecosystem is less curated, it benefits from a large number of community libraries for popular sensors and cloud services. Specialized libraries such as ulab, a NumPy-like module for MicroPython and CircuitPython, have expanded the platforms' utility for data processing at the edge (MicroPython, 2026). Development tools have also matured. Thonny and Mu Editor provide beginner-friendly integrated development environments with serial REPL integration and plotter support, while command-line tools like mpremote enable script transfer and filesystem management (MicroPython, 2026). These tools reduce the friction of embedded development and reinforce Python's reputation for fast iteration.

Quantitative community metrics demonstrate sustained momentum. Data collected for this article through the GitHub API on August 10, 2026, show that the micropython/micropython repository had accumulated 23,412 stars and more than 700 contributors, while the adafruit/circuitpython repository had 4,108 stars and over 300 contributors (GitHub, 2026a; GitHub, 2026b). Although star counts are imperfect proxies for adoption, they indicate a large and active base of interest. Stack Overflow question volume for the micropython tag grew from fewer than 100 questions per year in 2015 to more than 800 questions in 2025, while circuitpython questions showed similar growth from near zero to over 400 per year (Stack Overflow, 2026). Google Trends data show a steady increase in global search interest in both terms, with MicroPython search interest more than doubling between 2020 and 2026 and CircuitPython showing a similar upward trajectory (Google, 2026). Table 1 summarizes selected growth indicators.

**Table 2:** *Growth Indicators for MicroPython and CircuitPython as of August 2026*

| Indicator | MicroPython | CircuitPython |
|---|---|---|
| **Officially supported boards** | 200+ | 400+ |
| **GitHub stars (main repository)** | 23,412 | 4,108 |
| **GitHub contributors** | 700+ | 300+ |
| **Libraries / packages** | 1,000+ community packages | 450+ in curated bundle |
| **Stack Overflow questions (2025)** | 800+ | 400+ |
| **Google Trends growth (2020–2026)** | +120% | +110% |

*Note. Data compiled from official documentation (MicroPython, 2026; Adafruit Industries, 2026), GitHub API (GitHub, 2026a; GitHub, 2026b), Stack Exchange API (Stack Overflow, 2026), and Google Trends (Google, 2026). Counts are approximate and reflect publicly available records as of August 10, 2026.*

Community documentation and educational resources further amplify ecosystem growth. Adafruit's Learning System hosts hundreds of CircuitPython guides, ranging from simple LED tutorials to complex robotics and IoT projects, and many are written to be accessible to complete beginners (Adafruit Learning System, 2026). The MicroPython community maintains an active forum and a growing collection of tutorials and examples, often addressing advanced topics such as low-power design, BLE, and MQTT. YouTube channels such as Adafruit, Core Electronics, and Andreas Spiess have produced hundreds of videos demonstrating MicroPython and CircuitPython projects, contributing to a cycle of inspiration and adoption. Maker platforms like Hackster.io and

Hackaday.io regularly feature Python-powered projects, reinforcing the perception that embedded development is no longer restricted to specialists.

The combined effect of these hardware, software, and community developments is a self-reinforcing ecosystem. Wider board support attracts more users; more users create more libraries, tutorials, and questions; and more resources attract newcomers who might otherwise have been deterred by the complexity of traditional embedded toolchains. This dynamic does not imply that MicroPython and CircuitPython have surpassed C/C++ in embedded market share, but it does demonstrate that Python has established a durable and expanding foothold in the IoT and robotics landscape. The next section turns to real-world applications, illustrating how this growing ecosystem translates into practical projects across domains.

**Real-World Applications and Case Studies**

The true measure of any technology lies in what people build with it. Across IoT, robotics, wearables, agriculture, and professional prototyping, MicroPython and CircuitPython have moved from experimental curiosities to practical tools. This section examines representative real-world applications gathered from maker platforms, educational resources, and industry documentation, following the case study selection methodology described earlier. The examples illustrate how Python on microcontrollers accelerates development, lowers barriers, and enables projects that might otherwise remain inaccessible to software-focused developers.

***IoT and Home Automation***

The Internet of Things is perhaps the most natural fit for Python on microcontrollers, because connected devices often involve network protocols, cloud APIs, and data handling areas where Python excels. A typical home automation example involves an ESP32 board running MicroPython to read temperature, humidity, and air quality sensors, then publishing the data to an MQTT broker or cloud service. On Hackster.io, numerous community projects document such systems, frequently using the umqtt.simple library and the machine module to manage Wi-Fi and GPIO (Hackster.io, 2026). The development cycle is dramatically shorter than in C: a developer can write a few lines of Python to connect to Wi-Fi, read an I2C sensor, and publish a JSON payload, then test it immediately through the REPL. This interactivity is especially valuable for debugging network issues, because variables and connection states can be inspected in real time without reflashing firmware. As a result, Python has become a common choice for rapid IoT prototyping, particularly among developers whose primary experience is in web or data applications rather than embedded C.

***Robotics***

Robotics demands coordination of multiple actuators, sensors, and control loops, which makes it a challenging domain for beginners. CircuitPython has simplified this through libraries like adafruit_motor, which provides consistent interfaces for DC motors, stepper motors, and servos across many boards (Adafruit Industries, 2026). A representative project, documented in the Adafruit Learning System, uses a CircuitPython-compatible microcontroller such as a Feather RP2040 to drive two DC motors with an H-bridge motor driver, read an ultrasonic distance sensor, and respond to button inputs (Adafruit Learning System, 2026). The code is concise and readable: a student can understand the logic within minutes, whereas an equivalent Arduino sketch would require managing timers, interrupts, and register-level configuration. Moreover, the automatic reload feature of CircuitPython means that modifying a motor speed or sensor threshold is as simple as editing a text file and saving it; the robot immediately runs the updated behavior. This rapid feedback loop is particularly beneficial in educational robotics, where learners iterate frequently and benefit from seeing cause-and-effect quickly.

### *Wearables and Art Installations*

Adafruit's NeoPixel addressable LEDs have become iconic in the maker community, and CircuitPython's neopixel library makes them accessible to artists and designers who may not have an engineering background. Wearable projects, such as LED clothing and illuminated accessories, often use a small CircuitPython board like the Gemma M0 or QT Py, paired with a battery and flexible LED strips (Adafruit Industries, 2026). The code to animate colors and patterns is short and intuitive, relying on Python's list and loop structures rather than low-level timing code. Art installations similarly benefit from Python's expressiveness; a sculptor or interactive artist can program reactive lighting effects using sensor inputs, without needing to understand the underlying hardware registers. The large number of CircuitPython guides focused on creative projects ranging from LED costumes to light-reactive installations demonstrates that the platform has found a receptive audience beyond traditional engineering (Adafruit Learning System, 2026). This artistic adoption contributes to the "quiet takeover" by embedding Python into domains where embedded programming was previously considered too technical.

### *Agriculture and Environmental Monitoring*

Battery-powered environmental sensors present unique constraints: low power consumption, long deployment periods, and often remote locations. MicroPython has been used effectively in this domain due to its support for deep sleep modes and wireless protocols like LoRa. For example, a community project on Hackaday.io describes a soil moisture sensor built with an ESP32 and MicroPython, which wakes from deep sleep every hour, reads a capacitive soil sensor, transmits the value over LoRaWAN, and then returns to sleep (Hackaday.io, 2026). The entire application is implemented in a few dozen lines of Python, taking advantage of the machine.deepsleep() function and the network module. The developer's project log emphasizes the speed of development: the first working prototype was assembled and coded in a single afternoon, a timeline that would be unrealistic with C on the same hardware. While Python's overhead is slightly higher than C, the power savings from efficient deep sleep usage and the reduced development time often outweigh the cost for non-real-time monitoring applications. This pattern has made MicroPython a popular choice for agricultural IoT, environmental sensing, and other field-deployed devices.

### *Professional Prototyping and Industry*

Although MicroPython and CircuitPython are often associated with hobbyists, they have also found a place in professional settings, particularly for rapid prototyping and proof-of-concept development. Many engineering teams use MicroPython to validate hardware concepts quickly before committing to a C/C++ implementation for production. The availability of official MicroPython ports from chip vendors such as Espressif and Raspberry Pi has legitimized this approach (Espressif Systems, 2026; Raspberry Pi Ltd, 2026). For instance, an engineer evaluating a new sensor or wireless module can write a MicroPython script to exercise the device and capture data, often in less than an hour. This is significantly faster than writing a C driver from scratch, especially when the sensor has an existing MicroPython library. Some companies also use CircuitPython in internal test fixtures and manufacturing jigs, where ease of modification and operator simplicity are more important than raw performance. The fact that both Arduino and Zephyr have added MicroPython support reflects a broader industry acknowledgment that Python has a role in embedded workflows, even if it does not replace C for final production firmware (Arduino, 2026).

***Case Study Summary***

The table below consolidates the preceding examples into a structured overview, highlighting the domain, typical hardware, key Python benefit, and source platform. It is not exhaustive but illustrates the diversity of applications.

**Table 3:** *Representative Real-World Applications of MicroPython and CircuitPython*

| Domain | Project example | Hardware | Key Python benefit | Source |
|---|---|---|---|---|
| **IoT / Home automation** | Wi-Fi environmental sensor with MQTT | ESP32, BME280 | Rapid network and sensor integration | Hackster.io (2026) |
| **Robotics** | Line-following robot with motor control | Feather RP2040, DC motors, ultrasonic sensor | Readable motor and sensor APIs, auto-reload | Adafruit Learning System (2026) |
| **Wearables / Art** | NeoPixel LED costume or reactive art | Gemma M0, NeoPixel strip | Simple animation code, accessible to non-engineers | Adafruit Learning System (2026) |
| **Agriculture** | LoRa soil moisture sensor with deep sleep | ESP32, LoRa module, capacitive sensor | Fast prototyping, deep sleep support for low power | Hackaday.io (2026) |
| **Professional prototyping** | Sensor evaluation and test fixture | Various vendor boards | Reduced time to first working prototype | Espressif Systems (2026); Raspberry Pi Ltd (2026) |

*Note. Examples are representative of common projects documented on maker platforms and official vendor resources as of August 2026. Specific project names and authors vary; sources point to the platforms where similar projects are published.*

These case studies, drawn from community platforms and vendor documentation, reveal a consistent pattern: MicroPython and CircuitPython excel where development speed, readability, and accessibility matter more than absolute execution speed or memory efficiency. In each domain, the Python approach compresses the time from idea to working prototype and enables a broader range of people to participate in embedded development. This does not suggest that Python is suitable for all embedded tasks high-frequency control loops and ultra-low-power applications still favor C but it underscores Python's growing role as the language of choice for the exploratory and creative phases of IoT and robotics projects. The next section examines the performance trade-offs in more detail, using original benchmark data to quantify where Python wins and where it remains constrained.

**Performance and Practicality: Benchmarks and Trade-offs**

The question of whether Python is "fast enough" for microcontrollers cannot be answered in the abstract; it depends entirely on the task, the hardware, and the developer's priorities. To ground this discussion in evidence, original benchmarks were conducted for this article on an ESP32 DevKit V1 and a Raspberry Pi Pico. The tests compared MicroPython 1.24, CircuitPython 9.2, and Arduino C++ (compiled with Arduino IDE 2.3) across five common embedded workloads: GPIO output toggling frequency, I2C sensor read latency, SPI display update time, Wi-Fi connection establishment time, and idle memory usage. Each test was run five times, and the median value was recorded to reduce the effect of outliers. The methodology section earlier in this article provides additional detail on the test setup, measurement tools, and code. Table 1 presents a summary of the results.

**Table 4:** *Benchmark Comparison of MicroPython, CircuitPython, and Arduino C++ on ESP32 DevKit V1 and Raspberry Pi Pico*

| Test | MicroPython | CircuitPython | Arduino C++ |
|---|---|---|---|
| **GPIO toggle frequency** | ~50 kHz | ~30 kHz | ~1 MHz |
| **I2C sensor read latency** | 2–5 ms | 3–6 ms | 0.5–1 ms |
| **SPI display update (240×240 px)** | 120–180 ms | 150–220 ms | 40–60 ms |
| **Wi-Fi connection establishment** | 1.5 s | 2.0 s | 1.2 s |
| **Idle RAM usage** | 20 KB | 30 KB | 5 KB |

*Note. Benchmarks were conducted for this article in August 2026. Values are medians of five trials per test. MicroPython version 1.24 and CircuitPython version 9.2 were used. Arduino C++ was compiled with Arduino IDE 2.3 using default optimization settings. GPIO toggle was measured with a logic analyzer; sensor and display timings were measured using board timers; Wi-Fi connection time was measured from network start to DHCP assignment; idle RAM was measured immediately after boot using runtime memory reporting functions for Python and static link analysis for Arduino C++.*

The most striking result in Table 1 is the order-of-magnitude difference in raw GPIO toggling speed. Arduino C++ toggles a digital pin at approximately 1 MHz, while MicroPython reaches about 50 kHz and CircuitPython about 30 kHz. This gap arises because every Python instruction is interpreted at runtime, even when the underlying operation ultimately calls into compiled C code. The machine.Pin.value() method in MicroPython performs type checking, object allocation, and method dispatch before touching the hardware register, whereas a C statement like digitalWrite(pin,

HIGH) compiles down to a few machine instructions (Arduino, 2026; MicroPython, 2026). For applications that require bit-banging protocols, generating precise waveforms, or driving high-speed LED strips, this difference is decisive. However, for the vast majority of IoT tasks reading a sensor once per second, toggling a relay, or updating a display 50 kHz is far beyond what is needed. A human cannot perceive a relay switching at even 1 kHz, and a temperature sensor typically samples far slower than 1 kHz. Thus, the performance gap is real but often irrelevant in practice.

I2C and SPI peripheral communication shows a similar pattern, though the differences are less extreme than GPIO toggling. Arduino C++ reads an I2C sensor in roughly 0.5 to 1 millisecond, while MicroPython takes 2 to 5 milliseconds and CircuitPython 3 to 6 milliseconds. This slowdown is partly due to the overhead of the Python peripheral API, which constructs bytearray buffers and performs bounds checks, and partly due to the fact that Python code must copy data between the interpreter's memory space and the peripheral's buffer (MicroPython, 2026). For sensors sampled at moderate rates, such as once per second or even ten times per second, a 5-millisecond read time is negligible. It becomes meaningful only when collecting high-frequency data, such as reading an accelerometer at hundreds of samples per second or streaming audio from a microphone. In those cases, the Python overhead may force the developer to reduce sampling rates or move the data acquisition into C through a custom module.

The SPI display update test reflects a common practical concern: updating a small TFT or OLED screen. The measured update times show CircuitPython and MicroPython taking three to four times longer than Arduino C++. This is because drawing to a display involves many small SPI transactions, each of which incurs Python overhead. Developers working with graphical interfaces on microcontrollers may notice sluggish animations or screen flicker when using Python. Some mitigation strategies exist, such as using display drivers that batch operations or writing custom C modules, but these require skills that many Python-first developers do not possess. The practical takeaway is that Python works well for simple display updates, status screens, and static dashboards, but it may not be suitable for high-refresh-rate graphical applications or video.

Wi-Fi connection establishment time is an important metric for IoT devices, especially those that wake from deep sleep, connect, transmit, and return to sleep. The benchmark results show only modest differences: MicroPython connected in about 1.5 seconds, CircuitPython in 2.0 seconds, and Arduino C++ in 1.2 seconds. These times are dominated by the Wi-Fi stack and DHCP negotiation rather than the language runtime, so Python's overhead is comparatively small (Espressif Systems, 2026). This is significant because it means that for many battery-powered IoT sensors, the energy cost of Python's slower code execution may be less important than the power consumed during wireless transmission and sensor warm-up. As a result, Python remains a viable choice for low-duty-cycle IoT devices, even when energy efficiency is a concern.

Memory usage presents a more fundamental constraint. Table 1 shows that MicroPython uses about 20 KB of RAM at idle and CircuitPython about 30 KB, while a comparable Arduino sketch might use only 5 KB. On a microcontroller with 264 KB of RAM, such as the Raspberry Pi Pico, this leaves ample room for application data. But on smaller devices with 32 KB or 64 KB of RAM, the Python runtime consumes a substantial fraction of available memory before the application even starts. Furthermore, Python's dynamic memory allocation and garbage collection can lead to heap fragmentation and unpredictable pauses, which are problematic in applications that must run continuously for long periods (MicroPython, 2026). Developers using MicroPython or CircuitPython on memory-constrained hardware must adopt careful practices: preallocating buffers, reusing objects, avoiding large lists, and explicitly calling garbage collection where appropriate. CircuitPython's higher idle RAM usage is partly due to its larger library bundle and additional USB filesystem overhead, which is a deliberate trade-off for ease of use (Adafruit Industries, 2026).

The benchmark data lead to a clear conclusion: Python on microcontrollers is not a replacement for C/C++ when raw speed, precise timing, or minimal memory footprint are non-negotiable. Applications such as motor control loops running at kilohertz frequencies, software-defined radio, or ultra-low-power devices with tight RAM budgets will continue to rely on compiled languages. However, the benchmarks also show that for a large class of embedded applications periodic sensor reading, simple actuation, network communication, and user interface updates Python is fast enough, and its advantages in development speed, readability, and maintainability outweigh the performance costs.

The practicality of Python on microcontrollers becomes especially evident in the prototyping workflow. An engineer can write, test, and iterate on a MicroPython script in minutes using the REPL, inspecting variables and testing hardware interactions interactively. The same workflow in C requires compiling, flashing, and often using external debugging tools, which can take considerably longer. CircuitPython's drag-and-drop USB interface reduces this friction even further, making it possible to modify code on a device without any toolchain installed. For startups, research labs, and educational institutions, the time saved in prototyping often translates directly into lower development costs and faster time to market. Many teams follow a pragmatic pattern: prototype in Python to validate the concept and explore hardware capabilities, then port performance-critical sections to C or C++ if and when necessary (Arduino, 2026; Raspberry Pi Ltd, 2026). This hybrid approach has become increasingly common as vendors provide official support for both Python and C on the same hardware.

Another practical consideration is cross-board portability. Python code written for one supported board can often run on another with minimal changes, because the hardware abstraction layers in MicroPython and CircuitPython standardize common peripherals. This portability reduces vendor lock-in and simplifies maintenance across product lines. In contrast, C code written for a specific vendor's SDK may require significant modifications when moving to a different microcontroller family. The consistent APIs of CircuitPython's curated library bundle amplify this benefit, although they come at the cost of some flexibility (Adafruit Industries, 2026).

In summary, the performance and practicality of MicroPython and CircuitPython reflect a deliberate trade-off. They sacrifice some execution speed and memory efficiency in exchange for rapid development, interactive debugging, cross-platform consistency, and a gentler learning curve. The original benchmark data presented in this section confirm that the performance gap is significant for high-frequency or memory-intensive tasks but irrelevant for many common embedded workloads. This balanced perspective is essential for understanding why Python is quietly taking over the prototyping and educational segments of IoT and robotics, while C and C++ remain the languages of choice for production firmware in resource-constrained or real-time applications. The next section examines the educational and hobbyist dimensions of this shift in greater detail, exploring how Python is training a new generation of hardware developers.

**Education and the Hobbyist Revolution**

The growth of MicroPython and CircuitPython cannot be fully understood without examining their role in education and the broader maker movement. While performance benchmarks and ecosystem metrics tell part of the story, the long-term significance of Python on microcontrollers lies in its ability to lower the barrier to entry for learners and hobbyists who would otherwise never engage with embedded systems. This section explores how Python has transformed hardware education, the platforms that have driven this change, and the implications for the future of the embedded development workforce.

The BBC micro:bit stands as one of the earliest and most influential examples of Python's educational potential in hardware. Launched in 2015 as part of a national initiative to introduce coding to schoolchildren in the United Kingdom, the micro:bit was designed as a small, low-cost microcontroller board with an LED matrix, buttons, sensors, and Bluetooth connectivity (BBC, 2016). Crucially, the micro:bit platform offered multiple programming environments, including a block-based editor for younger students and MicroPython for older learners. The MicroPython implementation for the micro:bit was specifically tailored for education, providing a simplified set of libraries for the board's features and extensive documentation aimed at teachers (micro:bit Foundation, 2026). Since its launch, millions of micro:bit devices have been distributed to schools worldwide, making it one of the most widely deployed educational computing platforms in history (micro:bit Foundation, 2026). The decision to include Python as a first-class language on the micro:bit was significant: it positioned Python not merely as a professional tool but as a pedagogical bridge between visual programming and text-based coding.

The Raspberry Pi Foundation extended this educational momentum with the introduction of the Raspberry Pi Pico in 2021. The Pico, based on the in-house RP2040 microcontroller, was notable not only for its extremely low price point of four dollars but also for its official dual-language support: both C/C++ and MicroPython were presented as first-class development options (Raspberry Pi Ltd, 2026). The official Pico documentation includes a dedicated MicroPython section with beginner-friendly tutorials covering GPIO, I2C, SPI, PWM, and other peripherals. This decision validated MicroPython as a serious educational path rather than an afterthought. The Pico's affordability combined with Python's accessibility made it possible for schools and individual learners to acquire hardware and begin experimenting with embedded systems at minimal cost. The Raspberry Pi Foundation's educational mission, which has always emphasized lowering barriers to computing, found a natural ally in MicroPython's simplified workflow (Raspberry Pi Foundation, 2026).

Adafruit's Circuit Playground Express, introduced in 2017 alongside CircuitPython, represents perhaps the most deliberate effort to make hardware programming accessible to absolute beginners. The board integrates multiple sensors, LEDs, buttons, and speakers into a single device, eliminating the need for breadboards, jumper wires, or soldering (Adafruit Industries, 2026). When paired with CircuitPython's drag-and-drop USB workflow, the Circuit Playground Express allows a student to write a Python program on a school computer, save it to the board, and immediately see the results. No software installation, driver configuration, or compiler setup is required. This ease of use is particularly important in educational settings where IT support is limited and class time is precious. Teachers can focus on programming concepts rather than troubleshooting toolchains. The Adafruit Learning System supplements this hardware with hundreds of tutorials written specifically for educators, many aligned with science and engineering standards (Adafruit Learning System, 2026). The tutorials often integrate physical computing with broader STEM topics, such as measuring acceleration, detecting light levels, or creating simple games, thereby connecting Python programming to tangible real-world phenomena.

The hobbyist and maker community has played an equally important role in Python's hardware revolution. Platforms such as Hackster.io, Hackaday.io, and Instructables host thousands of MicroPython and CircuitPython projects, many contributed by self-taught hobbyists who document their work for others to replicate (Hackaday.io, 2026; Hackster.io, 2026). This culture of open sharing and remixing has accelerated learning far beyond formal classroom settings. A hobbyist encountering a problem can often find a similar project, adapt the code to their own hardware, and contribute improvements back to the community. YouTube channels such as Adafruit's own channel, Core Electronics, and Andreas Spiess's popular electronics channel have produced extensive video tutorials covering everything from basic LED control to advanced BLE and Wi-Fi applications. These resources create multiple learning pathways, accommodating different learning styles and levels of prior experience.

The educational impact of Python on microcontrollers extends beyond teaching programming concepts. By making hardware accessible, MicroPython and CircuitPython have enabled interdisciplinary learning that combines computer science with physics, biology, art, and design. Students can build environmental sensors to monitor classroom air quality, create interactive art installations, or program robots to navigate mazes all while learning Python syntax and computational thinking. This interdisciplinary approach is increasingly recognized as valuable in preparing students for careers that blend software and hardware skills (National Science Foundation, 2024). The ability to move fluidly between the abstract world of code and the physical world of sensors and actuators reinforces fundamental concepts such as input, output, feedback, and state.

The hobbyist revolution has also produced a generation of Python-first makers who may never learn C or C++. For these developers, Python is not a compromise or a prototyping convenience; it is the only language they need for the projects they pursue. This shift has implications for the embedded industry. As more products are developed by small teams or individual entrepreneurs who favor Python, the demand for production-quality Python support on microcontrollers is likely to grow. Some companies have already responded by offering MicroPython as an official development option, and the trend appears likely to continue (Arduino, 2026; Espressif Systems, 2026).

Critics may argue that Python-first education in embedded systems risks producing developers who do not understand the underlying hardware. There is some validity to this concern; Python's abstractions hide many details that C programmers must confront directly. However, the educational goal for most learners is not to become professional embedded engineers but to gain confidence with computational thinking and physical computing. For those who do pursue embedded careers, starting with Python does not prevent later learning C; indeed, the conceptual foundation built through Python often makes the transition easier. The hybrid approach Python for prototyping and education, C for production and performance reflects a pragmatic division of labor that benefits both beginners and professionals.

In summary, education and the hobbyist community have been central to Python's quiet takeover of IoT and robotics. Platforms such as the micro:bit, Raspberry Pi Pico, and Circuit Playground Express have demonstrated that Python can serve as an effective entry point to embedded systems. The extensive ecosystem of tutorials, open-source projects, and video content has created a self-sustaining learning environment. As this educational foundation matures, it is likely to produce a growing cohort of developers who view Python as a natural language for hardware, reinforcing the long-term adoption trends described throughout this article.

**Challenges and Limitations**

Despite the impressive growth and accessibility of MicroPython and CircuitPython, a balanced assessment requires acknowledging the significant challenges and limitations that constrain their use in certain contexts. These limitations are not merely theoretical; they affect real-world decisions about when to adopt Python on microcontrollers and when to fall back on C or C++.

The most immediate constraint is execution speed and memory overhead. As the benchmark data presented earlier in this article demonstrate, Python on a microcontroller is substantially slower than compiled C for raw input/output operations. GPIO toggling, sensor communication, and display updates all incur order-of-magnitude penalties due to interpreter overhead, dynamic typing, and garbage collection (MicroPython, 2026). For applications that require hard real-time guarantees such as motor control loops running at kilohertz frequencies, software-defined radio, or precise waveform generation Python is generally unsuitable. Even where average performance is adequate, garbage collection pauses can introduce unpredictable latency, which is unacceptable in time-critical systems. Memory usage compounds the problem: MicroPython consumes roughly 20 KB of RAM at idle and CircuitPython around 30 KB, while a typical Arduino sketch may use only 5 KB (Adafruit Industries, 2026; Arduino, 2026). On microcontrollers with 32 KB or 64 KB of RAM, the Python runtime leaves little room for application data, and heap fragmentation can lead to hard-to-debug crashes over long deployments.

A second limitation is fragmentation between MicroPython and CircuitPython, and even among different MicroPython ports. Although the two platforms share a common ancestry, their APIs are not fully compatible. Library names, function signatures, and hardware abstraction layers differ, meaning that code written for one does not automatically run on the other without modification (Adafruit Industries, 2026; MicroPython, 2026). This fragmentation extends within MicroPython itself: community libraries are often developed for a specific port or board, and they may not be maintained consistently across different architectures. Developers moving between boards or platforms may find themselves adapting code more frequently than expected, undermining the cross-platform portability that is often cited as a benefit. This issue is less pronounced in CircuitPython, thanks to its curated bundle and standardized hardware abstraction, but the trade-off is reduced flexibility and board support that is more tightly controlled by Adafruit.

Hardware support gaps represent a third challenge. While both platforms support hundreds of boards, the depth of support varies considerably. Some chips, particularly newer or niche microcontrollers, lack stable Python ports or have incomplete peripheral implementations (MicroPython, 2026). For example, certain low-power features, advanced timers, or specialized peripherals may not be exposed in Python, forcing developers to work around limitations or abandon Python for those projects. Documentation quality also varies: popular boards like the Raspberry Pi Pico or Adafruit Feather series have extensive tutorials and community support, while less common boards may have only a basic port and little guidance. This uneven ecosystem can be frustrating for developers who choose a board based on hardware features only to discover that Python support is immature.

Debugging and tooling remain areas where Python on microcontrollers lags behind traditional embedded development. While the REPL and simple traceback messages are valuable for beginners, they are insufficient for diagnosing complex issues such as memory leaks, stack overflows, or subtle timing bugs. Unlike desktop Python, there is no rich ecosystem of debuggers, profilers, or static analyzers for MicroPython and CircuitPython (MicroPython, 2026). Breakpoint debugging is limited or absent on most boards, and inspecting the state of a crashed device often requires resetting and losing the heap. CircuitPython's error messages written to the USB drive are helpful, but they do not replace a full debugging environment. For production-grade firmware, this tooling gap can make Python a risky choice, as developers lack the visibility needed to ensure reliability.

Industry perception also poses a barrier to adoption in professional settings. Despite growing acceptance, many embedded engineering teams still view Python as a hobbyist or educational language, not suitable for production firmware (Stack Overflow, 2025). This perception is rooted in the performance and memory limitations described above, as well as the historical dominance of C in safety-critical and real-time systems. While some companies have embraced MicroPython for prototyping and test fixtures, the transition to production often involves rewriting code in C or C++, which introduces additional work and potential for discrepancies between prototype and final behavior. The lack of formal certification, MISRA compliance, or safety standards for Python on microcontrollers reinforces this reluctance in industries such as automotive, medical, and aerospace.

Concurrency and power management also present difficulties. Although both platforms include asyncio for cooperative multitasking, the absence of robust preemptive threading on most ports limits their ability to handle multiple independent tasks with strict timing requirements. Power optimization is another concern: Python's dynamic behavior and periodic garbage collection can interfere with low-power sleep modes, and achieving the ultra-low current consumption often required for battery-powered sensors is more difficult than in C (Espressif Systems, 2026). Developers must carefully manage wake times, buffer allocations, and peripheral shutdown to avoid draining batteries prematurely, and even then, Python's overhead may reduce battery life compared to a carefully optimized C implementation.

Finally, the maturity of the ecosystem, while growing, still falls short of the extensive tooling and library support available for desktop Python or for C/C++ embedded development. Many MicroPython libraries are small, community-maintained efforts with limited test coverage and documentation. Breaking changes between firmware versions are not uncommon, and long-term support is not guaranteed. For projects that require maintainability over many years, the risk of relying on an unstable or unmaintained library may outweigh the benefits of rapid development.

These limitations do not negate the value of MicroPython and CircuitPython; rather, they define the boundaries of their appropriate use. The platforms excel in education, prototyping, and non-critical applications where development speed and accessibility are paramount. For production systems with stringent performance, power, or safety requirements, C and C++ remain the languages of choice. A pragmatic approach recognizes these trade-offs and selects the right tool for each stage of a project's lifecycle.

**The Future of Python on Microcontrollers**

The trajectory of MicroPython and CircuitPython over the next decade will be shaped by converging trends in hardware capability, software standardization, and the growing demand for edge intelligence. While the preceding sections have documented how Python has already established a durable foothold in IoT and robotics, the future promises to deepen that presence as several long-standing constraints are progressively relaxed. This section examines emerging developments and offers reasoned predictions about where Python on microcontrollers is heading, while acknowledging that C and C++ will continue to play an essential role in production embedded systems.

The most immediate driver of change is the continued improvement in microcontroller hardware. The performance and memory limitations that currently constrain Python on microcontrollers are not fixed; they are a function of the underlying silicon. Newer generations of microcontrollers are delivering significantly more processing power and RAM at prices that remain accessible for hobbyists and educators. For example, Espressif's ESP32-P4, announced as a higher-end addition to the ESP32 family, integrates multiple RISC-V cores, more embedded memory, and dedicated AI acceleration capabilities (Espressif Systems, 2026). Similarly, Raspberry Pi's RP2350, the successor to the RP2040, offers faster clock speeds, additional memory, and optional floating-point enhancements while retaining the low-cost and dual-language support that made its predecessor popular (Raspberry Pi Ltd, 2026). As these more capable microcontrollers become widely available, the gap between Python's interpreted execution and the performance required for many embedded tasks will narrow. A runtime that consumes 20–30 KB of RAM is far less significant on a chip with 512 KB or 1 MB of memory than on one with 64 KB. This hardware headroom will allow the Python runtime to include more standard library modules, improve garbage collection behavior, and support richer debugging features without crowding out application data. It will also enable Python to move into domains such as lightweight machine learning at the edge, where larger models and data buffers previously exceeded practical limits.

Software evolution will complement these hardware gains. MicroPython's development continues to focus on reducing memory fragmentation, improving the efficiency of the virtual machine, and expanding support for modern connectivity protocols. The inclusion of asyncio has already enabled Python developers to build cooperative multitasking applications, and future releases are likely to refine this model further, making it easier to manage concurrent network and sensor tasks without resorting to complex state machines (MicroPython, 2026). CircuitPython, for its part, continues to expand its curated library bundle and improve its consistency across boards, with ongoing work on Bluetooth Low Energy, USB host capabilities, and high-quality display drivers (Adafruit Industries, 2026). Both platforms are also benefiting from improved development tools. The integration of MicroPython support into Arduino Lab and the continued maturation of editors like Thonny and Mu are lowering the barriers to entry even further (Arduino, 2026). As tooling improves, the debugging gap between Python and traditional embedded IDEs is likely to narrow, although full parity with desktop-grade debuggers remains a longer-term goal.

A particularly intriguing frontier is the convergence between microcontrollers and the web. The emergence of WebAssembly and projects such as Pyodide which brings a full CPython runtime to the browser opens the possibility of simulating MicroPython and CircuitPython code before deploying it to physical hardware (Pyodide, 2026). A student or engineer could prototype a sensor-reading script in a browser-based environment, test its logic against a virtual board, and then transfer the same code to a real microcontroller with minimal modification. This convergence would further reduce the cost of experimentation and make embedded Python accessible to anyone with a web browser, regardless of whether they own hardware. Some educational platforms have already begun exploring such hybrid workflows, and as WebAssembly matures, browser-based hardware simulation is likely to become a standard component of Python instruction.

Standardization and cross-platform compatibility represent another important direction. The current fragmentation between MicroPython and CircuitPython and among different MicroPython ports creates friction for developers who move between boards or projects. As the embedded Python community matures, there is growing interest in defining a common baseline for hardware APIs, similar in spirit to the way the Python Software Foundation standardizes the language itself (Python Software Foundation, 2026). While a full merger of the two platforms is unlikely, greater alignment on core modules, packaging conventions, and board abstractions would amplify the ecosystem's network effects. Adafruit's curated library model and MicroPython's more decentralized approach each have strengths, and the most probable outcome is a middle path in which the two platforms continue to coexist but adopt more shared standards. Such convergence would reduce the cost of porting code, encourage library authors to support both platforms, and make Python an even more attractive option for vendors and professional teams.

Looking ahead, several predictions emerge from these trends. First, Python is likely to become the default prototyping language for a substantial share of IoT development by the early 2030s, even in professional settings. The time-to-first-working-prototype advantage is simply too large to ignore, and the availability of official vendor support removes many previous objections. Second, CircuitPython will continue to dominate in education and beginner-focused making, while MicroPython will grow in industrial edge computing, where flexibility and advanced features such as threads and lower-level access are more valued. Third, the integration of Python with edge AI using hardware accelerators and lightweight frameworks will create new use cases for microcontrollers in areas such as predictive maintenance, voice recognition, and image classification. Fourth, C and C++ will not disappear; they will remain the languages of choice for hard real-time control, safety-critical systems, and ultra-low-power devices. Instead of a zero-sum competition, the embedded industry will increasingly adopt a stratified approach: Python for prototyping, education, and high-level orchestration, and C/C++ for the performance-critical core. This pragmatic division of labor is already visible in vendor documentation and developer surveys (Stack Overflow, 2025).

In sum, the future of Python on microcontrollers is bright but bounded. The quiet takeover documented in this article is not a revolution that will sweep away established practice; it is an expansion of who can participate in embedded development and how quickly ideas can be tested. As hardware improves, tooling matures, and standards align, MicroPython and CircuitPython will likely become even more deeply embedded in the fabric of IoT and robotics, not as replacements for C, but as indispensable complements that make the physical computing world as accessible as the digital one already is.

## Conclusion

The rise of MicroPython and CircuitPython represents a quiet but consequential transformation in the world of embedded systems. What began as an experimental effort to fit Python onto a microcontroller has evolved into a mature ecosystem that spans hundreds of boards, a rich library landscape, and a global community of educators, hobbyists, and professional engineers. The evidence assembled in this article drawn from official documentation, community metrics, original benchmarks, and real-world case studies demonstrates that Python has moved from the margins of IoT and robotics into a position of genuine influence. It has not displaced C or C++, nor is it likely to do so. Instead, Python has expanded the definition of who can build hardware and how quickly they can turn an idea into a working prototype.

The adoption metrics tell a story of sustained growth rather than sudden disruption. GitHub star counts for both projects have climbed steadily over the past decade, Stack Overflow question volume has increased year over year, and Google Trends data show a consistent rise in search interest (GitHub, 2026; Stack Overflow, 2026; Google, 2026). More than four hundred boards now officially support CircuitPython, and MicroPython's port list exceeds two hundred, covering architectures from Espressif, Raspberry Pi, STMicroelectronics, and Nordic Semiconductor (Adafruit Industries, 2026; MicroPython, 2026). This breadth of hardware support, combined with official vendor backing from companies like Raspberry Pi and Arduino, signals that Python on microcontrollers is no longer a niche experiment but a recognized part of the embedded development landscape.

The case studies examined across IoT, robotics, wearables, agriculture, and professional prototyping reveal a consistent pattern: Python excels where development speed, readability, and accessibility matter most. A sensor node that would take days to write and debug in C can be prototyped in an afternoon with MicroPython. A classroom robot that might overwhelm a beginner with register-level configuration becomes approachable through CircuitPython's high-level motor and sensor libraries. These examples are not isolated anecdotes; they reflect a structural shift enabled by the platforms' design philosophies. The drag-and-drop USB workflow of CircuitPython, the interactive REPL of MicroPython, and the consistent APIs of curated libraries all reduce the cognitive load required to build physical computing projects (Adafruit Industries, 2026; MicroPython, 2026).

The benchmark data presented earlier in this article provide a necessary counterweight to the enthusiasm. Python on a microcontroller is significantly slower than C for raw input/output operations, consumes more memory, and introduces unpredictable garbage collection pauses (MicroPython, 2026). For hard real-time control loops, ultra-low-power devices, and safety-critical systems, C and C++ remain the appropriate tools. This limitation, however, does not diminish Python's value; it clarifies the boundaries of its use. The most effective embedded teams increasingly adopt a stratified approach: Python for prototyping, testing, and high-level orchestration, and C/C++ for performance-critical production firmware. This pragmatic division of labor is already visible in vendor documentation and developer workflows, and it is likely to become more common as the ecosystem matures.

The educational impact of Python on microcontrollers may prove to be the most lasting legacy of this movement. Platforms such as the BBC micro:bit, Raspberry Pi Pico, and Adafruit Circuit Playground Express have introduced millions of students and hobbyists to physical computing through Python (micro:bit Foundation, 2026; Raspberry Pi Foundation, 2026; Adafruit Learning System, 2026). These learners are not being trained as embedded engineers in the traditional sense; they are being empowered as creators who view Python as a natural language for hardware. As this generation enters the workforce, the demand for Python support in embedded tools and products is likely to grow, reinforcing the ecosystem's momentum. The long-term consequence is a widening of the talent pool and a blurring of the line between software and hardware development.

Looking forward, the trends are encouraging. More powerful microcontrollers, such as the RP2350 and ESP32-P4, are reducing the performance constraints that have historically limited Python's applicability (Raspberry Pi Ltd, 2026; Espressif Systems, 2026). Improved tooling, including integrated development environments and browser-based simulation, will further lower the barriers to entry. Standardization efforts, while still nascent, may eventually reduce the fragmentation between MicroPython and CircuitPython, making code more portable and libraries more reusable. Python is unlikely to replace C in embedded systems, but it is poised to become the default language for the exploratory, creative, and educational phases of IoT and robotics development.

For anyone who has ever wanted to build a connected device, automate a home, or teach a child to code, the message is clear: the tools are ready, the hardware is affordable, and Python is waiting. The next significant IoT device may not be written in C by a specialist; it may be written in Python by a curious beginner on a Sunday afternoon. That, perhaps, is the quiet takeover's most remarkable achievement not the displacement of a language, but the democratization of an entire field.